# Common Metrics for Analyzing, Developing and Managing Telecommunication Networks

Salman M. Al-Shehri[1], Pavel Loskot[1*], **Tolga Numanoğlu**[2], and Mehmet Mert[2]

## Abstract

The metrics play increasingly fundamental role in the design, development, deployment and operation of telecommunication systems. Despite their importance, the studies of metrics are usually limited to a narrow area or a well-defined objective. Our study aims to more broadly survey the metrics that are commonly used for analyzing, developing and managing telecommunication networks in order to facilitate understanding of the current metrics landscape. The metrics are simple abstractions of systems, and they directly influence how the systems are perceived by different stakeholders. However, defining and using metrics for telecommunication systems with ever increasing complexity is a complicated matter which has not been so far systematically and comprehensively considered in the literature. The common metrics sources are identified, and how the metrics are used and selected is discussed. The most commonly used metrics for telecommunication systems are categorized and presented as energy and power metrics, quality-of-service metrics, quality-of-experience metrics, security metrics, and reliability and resilience metrics. Finally, the research directions and recommendations how the metrics can evolve, and be defined and used more effectively are outlined.

## Index Terms

Energy metrics; quality of experience (QoE) metrics; quality of service (QoS) metrics; resilience and reliability metrics; security metrics; telecommunication networks.

[1] College of Engineering, Bay Campus, Swansea University, Swansea, SA1 8EN, United Kingdom
[2] ASELSAN A.S.-Communications and IT Division, Ankara, Turkey
[*] Corresponding author. E-mail: p.loskot@swan.ac.uk Tel.: +44 1792 602619 Fax: +44 1792 295676

# 1. INTRODUCTION

The metrics play a crucial role in the whole telecommunication system life cycle. They are used from the planning stages to specify key performance indicators (KPIs) to operational stages where they are included in service level agreements (SLAs) to monitor the system performance. The metrics are usually defined in different contexts to accomplish different objectives. For instance, the KPI metrics are usually selected to drive the system or service adoption rates by the customers whereas the SLA metrics are established to moderate contractual agreements between the service providers and the end-users. The standardized metrics enable comparison of similar products and services from different equipment manufacturers and service providers while also ensuring the compliance with the relevant governmental regulations. The metrics describe the properties of a physical infrastructure, or they characterize the dynamic processes occurring within such infrastructure. More generally, the metrics determine how the system under consideration is perceived by the various stakeholders. In this sense, the metrics are representing simple models (i.e., abstractions) of often very complex systems. However, as the complexity of man-made systems normally increases with time, the metrics may have to be evolved correspondingly in order to remain meaningful. For instance, the cellular networks in the late 90's were designed to mainly maximize their spectral efficiency. Such a simple design approach is no longer adequate to fulfill the today's requirements on the existing and the emerging cellular systems which are assuming multiple design and operational objectives and the corresponding metrics including reliability, spectral efficiency, energy consumption, and the capital and operational expenditures among others. In other cases, the metrics that were once used extensively across the telecommunication industries are no longer relevant such as the additional connections and the average revenue per user (ARPU) [1].

The complexity of telecommunication systems is largely driven by the increasing connectivity of networks, a plethora of applications and services offered, implementing more of the network functionality in software rather than hardware to enhance flexibility (e.g., to provide multitude of services over the shared infrastructure), making the networks more autonomous, adopting new business, computing and communication models, and by introduction of new technologies such as machine-to-machine (M2M) communications, the Internet of Things (IoT), and Cloud Computing. In addition, these systems are very dynamic as they adapt their configuration and operations to the external service demands in order to optimize the use of system resources.

It is clear that defining and understanding metrics is not straightforward, but it is a rather complex matter [2]. The fundamental importance of metrics, however, strongly motivates why this subject needs to be considered more seriously and more systematically than ever before. As the complexity of telecommunication systems continuous to grow, it is increasingly more difficult to select the right metrics to achieve the desired objectives, and to assist in operational decisions without overcomplicating the system design. Traditionally, the selection of metrics has been primarily driven by the experience and well defined design rules [3]. Later on, the benefits of standardizing some of the metrics were recognized, so the performance of different systems and components could be objectively compared. The standardized metrics prescribe not only what quantities are being measured or observed, but also under what system configuration as well as what any other conditions need to be satisfied. This is true, in general, for any kind of measurements: one has to define a complete set of assumptions that may influence these measurements. At the same time, it is important to exclude the assumptions which do not affect the measurements in order to avoid the misinterpretation of the metric. For example, the validity of assumptions which were once relevant to the measurements may have to be checked again after the system evolves and becomes more complex. Adopting the assumptions irrelevant to the measurements may lead to incorrect decisions, and cause severe security vulnerabilities.

In technical literature, many different metrics have been devised and adopted to study the systems. These metrics can be defined explicitly, or they are implicitly derived as a result of a mathematical analysis. Some of these metrics are used much more often than the others (e.g., the probability of outage, bit and frame error rate), so they may be considered to be informal standards for evaluating the system performances. However, there are also many other metrics defined and used in technical papers which did find such widespread adoption. These more rarely occurring metrics are usually specialized for particular system models which limit their broader adoption in more general cases.

A useful strategy for defining the meaningful metrics is to exploit the existing measured data, since the measurements represent some type of metrics themselves. This approach is particularly attractive in the recent deluge of Big Data. A subsequent data processing defines a target metric for observing the system, so many practical metrics can be created or defined hierarchically. The scenario where data measurements constrain possible applications and models of the system with the corresponding metrics is sometimes referred to as a reverse system modeling. This modeling strategy is also used when the metrics are defined empirically by the experiments. In contrast, we can define system models with their metrics for a given application, and collect only the data that are relevant. This approach is typical for engineering designs, and it is sometimes referred to as a forward system modeling. These two system modeling strategies to define the corresponding metrics are depicted in Fig. 1.

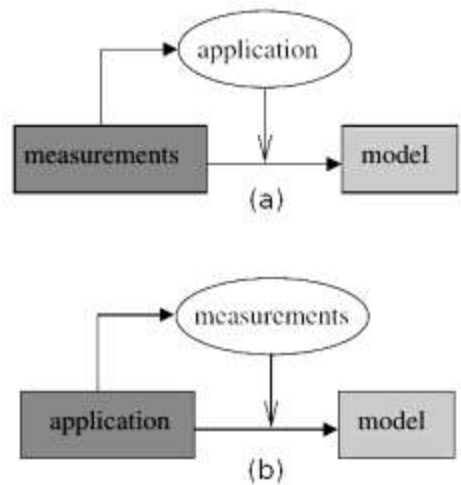

Fig. 1: The metrics definition strategies using reverse modeling (a), and forward modeling (b).

The following characteristics of the measurements can guide the selection of the corresponding metrics:

- accuracy: the measurement errors and biases need to be within acceptable limits;

- validity: the measurements and their evaluations need to be checked for correctness;

- feasibility: the measurements have to be collected as often as desired;

- robustness: the measurement quality must not be affected by changing conditions;

- efficiency: the measurements should not consume too much of the system resources;

- desirability: the measurements collected are required for the system design or operation;

- viability: the measurements being collected can clearly provide the measurable benefits.

The available literature about telecommunication metrics usually narrows down their focus to a specific category of metrics. For example, the survey [4] of common energy and power metrics in telecommunication systems does not explicitly distinguish between the metrics and the models of systems to quantify the energy, throughput and traffic load in the cellular networks. A generic metrics-driven framework for autonomous management of heterogeneous networks is given in [5]. This paper considers a system architecture and the corresponding reference models together with the key metrics for monitoring the telecommunication networks. The metrics in [6] are described and categorized according to their function; however, their definitions are not provided but referenced in other papers. A comprehensive study of fairness in telecommunication networks has been done in [7]. Both qualitative and quantitative fairness measures including their mathematical or other definitions are outlined for different protocol layers and for different network contexts.

More importantly, at present, there seem to be no established general theory how to define the optimum metrics for a given scenario. An equivalent task is to decide which of the two or more possible metrics is more optimal is some sense. Such theory would provide guidelines how to define new metrics, improve or modify the existing metrics, identify the metrics that need to be phased out and replaced, and decide how to bundle and weigh multiple metrics for the particular use cases. The optimum metrics are crucially dependent on who is using these metrics (different stakeholders have different goals and needs), and the assumptions and system models adopted for given applications and services provided by the system considered. Fig. 2 highlights the most important tasks and challenges when selecting the metrics for telecommunication systems.

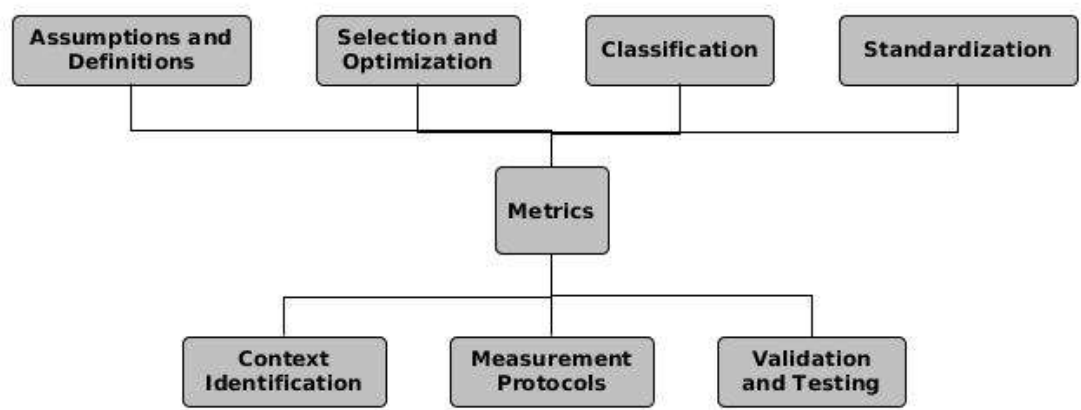

Fig. 2: The metrics: tasks and challenges.

The telecommunication network providers work with the equipment manufacturers to build and operate the network infrastructure. They are concerned with the KPIs and the cost related metrics such as a time-to-market, break-even point, return on investment (ROI) and the capital (CAPEX) and operational (OPEX) expenditures to account for the deployment (e.g. the site rental cost), installation and maintenance costs. These stakeholders pursue the energy efficiency in order to significantly reduce the OPEX, and they dimension the network capacity just ahead of the ever increasing demand. The telecommunication service and content providers rent the network infrastructure from the network providers to deliver the telecommunication services to the end-users who are sometime referred to as the service subscribers or consumers [8]. The service providers are concerned with the quality-of-service (QoS) and the quality-of-experience (QoE) metrics which are usually exemplified in the SLAs to drive the business operations. In addition, the service providers may assume the service uptime and the service availability, and other economic metrics such as the user churn rate. The telecommunication markets are governed by the government controlled regulatory bodies who are issuing the telecommunication policy frameworks, standards, and even the investment incentives. These institutions generally promote the end-user interests by balancing sometimes conflicting goals of the stakeholders, and by enforcing the telecommunication services to be accessible, affordable, secure and of acceptable quality [9].

The increasing complexity of telecommunication systems makes the classification or taxonomy of their metrics to be non-trivial. The complex systems in the industrial world such as the enterprises and their products are often efficiently described using so-called Zachman framework [10]. Since every metric represents a model of the system, this framework can be also used to classify the metrics of systems. In particular, Zachman framework allows a consistent and systematic modeling of complex systems from the perspective of different stakeholders as illustrated in Table I. The stakeholder perspectives are classified as the contextual, conceptual, logical, physical, and out-of-context views of the system, or equivalently, as the scope, business, system, technology, and detailed representation models of the system, respectively. These models are organized in a two-dimensional matrix with the stakeholder perspectives given in rows and the matrix columns represent the model attributes. Hence, the system model in each cell of the matrix can be used to define the corresponding metrics. Zachman framework for a telecommunication system may assume these model attributes:

- data model of the measurements (the what);

- flow of the measurements to make decisions (the how);

- locations of the measurements (the where);

- interaction with the decision makers (the who);

- time dependency of the measurements (the when);

- motivation for taking the measurements (the why).

In case of telecommunication systems, the perspectives of the following stakeholders should be considered: equipment manufacturers and subcontractors, infrastructure providers, service providers, content providers, network operators, end-users, governments, and regulatory bodies. Table II highlights different interests of the telecommunication stakeholders which, in general, may be conflicting. Therefore, the governments, sometimes through regulatory bodies, need to balance the stakeholder interests to deliver the telecommunication services in the desired quality and quantity across the whole region or a country for the benefit of the national and regional economies.

Table I
ZACHMAN FRAMEWORK OF A TELECOMMUNICATION ENTERPRISE

| | What (data) | How (function) | Where (network) | Who (people) | When (time) | Why (motivation) |
|---|---|---|---|---|---|---|
| Scope (contextual) | data models | data flows | coverage, locations | operators, vendors, end-users | network upgrades, expansions | objectives, strategies |
| Business (conceptual) | business relationships | business processes | business logistics | work flows | business events | business plan |
| System (logical) | data relationships | service architecture | distributed system | human computer interfaces | system events and cycles | business rules |
| Technology (physical) | measured data | system design | network architecture | system to user representation | action events | operation rules design |
| Detailed (out-of-context) | data definitions | algorithms | network protocols | jobs and responsibilities | time management | rules specifications |

Table II
STAKEHOLDER INTERESTS IN DIFFERENT ASPECTS OF A TELECOMMUNICATION SYSTEM

| | Energy consumption | Services | Costs |
|---|---|---|---|
| Equipment manufacturers | embodied energy | KPIs | production cost |
| Infrastructure providers and network operators | deployment and operational energy | SLAs | CAPEX and OPEX |
| Content and service providers | operational energy and battery life time | QoS and QoE | infrastructure rent cost |
| Regulatory bodies | spectral RF masks | KPIs and SLAs | fairness in customer fees |
| Government | total energy in the sector | availability and accessibility | GDP creation |
| End-users | battery life time | QoE | monthly fee |

A. CONTRIBUTIONS AND ORGANIZATION OF THE PAPER

General studies about the selection and the use of metrics and KPIs to design, operate and manage telecommunication systems and services appear to be still missing in the telecommunication literature. Hence, a tutorial style introductory section was devoted to this subject. The metrics were assumed to represent simple models or projections of systems. It was discussed that the optimum metrics selection is constrained either by the application being considered or by the available measurements. Even though one may define an optimum sensing strategy for given application and mathematical model of the system, finding the optimum system model may be an arduous task, particularly when the system is deemed to be complex. We also discussed the characteristics of measurements, and proposed to use Zachman framework for the enterprise modeling as a generic strategy for selecting the appropriate metrics and KPIs.

More importantly, our original intention was to summarize and classify all the metrics that are used in telecommunications network engineering. When trying to achieve this goal, we discovered two things. First, there are too many such metrics, so collecting all of them is impractical. Second, when we restricted our attention to only those metrics that are used the most frequently, several distinct categories of the metrics have readily emerged. We have then selected, in our opinion, the 5 most important categories of technical metrics to be surveyed in the following sections of this paper:

- Section 2: Energy and power metrics
- Section 3: Quality of service (QoS) metrics
- Section 4: Quality of experience (QoE) metrics
- Section 5: Security metrics
- Section 6: Reliability and resilience metrics

The last Section 7 concludes the paper, and provides some suggestions about the possible future research directions how to develop metrics for the emerging telecommunication systems.

Other major categories of metrics we decided not to include in this paper are the metrics related to the user behavior such as the metrics used in game theory, the metrics involving the user mobility, the metrics accounting for the social interactions [11] and fairness (e.g., of allocating the resources and access to services), and a large group of metrics which are used to assess the financial performance and business issues (e.g., return on investment, CAPEX, OPEX, and churn rate, to name a few). The metrics for evaluating the performance of communication systems at the lower layers of the protocol stack (e.g., probability of outage, spectral efficiency, and coverage at the physical layer, and round-trip delay, and throughput at the network layer) are generally considered under the umbrella of QoS metrics. The metrics pertinent to the application layer of the protocol stack (e.g., mean opinion score) are summarized as QoE metrics.

Our survey provides plentiful references to other more specialized papers on metrics. The other papers included in our references contain representative examples of the specific metrics. However, unlike the survey papers which are focusing on a certain type of metrics in some application domain (e.g., energy consumption metrics, and metrics for video quality assessment), our aim is to provide a much broader view on the metrics ecosystem in telecommunications engineering. This approach enables us to compare the developments of metrics in different metric categories, identify which types of metrics are used more frequently, and which metrics tend to become standardized. The readers can review what metrics are commonly used in telecommunications, and decide which of these metrics can be the most appropriate for their application, especially when a combination of several metrics is required or expected. Finally, we hope that this paper can stimulate further research into more systematic development of the new as well as modification and adaption of the existing telecommunication metrics as we will discuss in Section 7.

The following acronyms are used throughput this paper:

| | | | |
|---|---|---|---|
| 3GPP | 3rd Generation Partnership Project | NR | no reference |
| AQoS | application-based QoS | OPEX | operational expenditure |
| ARPU | average revenue per user | PASTA | "Poission arrivals see time averages" |
| CAC | call admission control | PEVQ | perceptual evaluation of video quality |
| CAPEX | capital expenditure | PSNR | peak signal-to-noise ratio |
| CIA | confidentiality, integrity and availability | QoE | quality of experience |
| ETSI | European Telecommun. Standards Institute | QoR | quality of recovery |
| FR | full reference | QoS | quality of service |
| ICMP | Internet Control Message Protocol | RFID | radio-frequency identification |
| ICT | Information and Commun. Technologies | ROI | return on investment |
| IEC | International Electrotechnical Commission | RR | reduced reference |
| IP | Internet Protocol | RTI | real-time interactive |
| ISO | Internat. Organization for Standardization | RTT | round-trip time |
| ISP | Internet service provider | SLA | service level agreement |
| ITU | International Telecommunication Union | SNMP | Simple Network Management Protocol |
| IoT | Internet of Things | SNR | signal-to-noise ratio |
| KPI | key performance indicator | SSIM | structural similarity index |
| LAN | local area network | TCB | trusted computing base |
| M2M | machine-to-machine | TCP | transmission control protocol |
| MANET | mobile ad-hoc network | UDP | user datagram protocol |

| MOS | mean opinion score | UE | user equipment |
|------|------|------|------|
| MSE | mean squared error | UX | user experience |
| MTBF | mean time between failures | VQM | video quality metric |
| MTTF | mean time to failure | VoD | video on demand |
| MTTR | mean time to repair | VoIP | voice over IP |
| NFV | network function virtualization | WSN | wireless sensor network |
| NIST | National Inst. of Standards and Technology | WiMAX | Worldwide Interoper. for Microwave Access |
| NQoS | network-based QoS | ZING | client-service application layer protocol |

## 2.  ENERGY AND POWER METRICS

Using the energy efficient equipment is driven by the need to lower the operational cost, extend the operational life time of handheld, battery powered devices, and to reduce the environmental impact. However, the reduction of the energy consumption is constrained by maintaining the required quality of offered telecommunication services. The energy is an integrative quantity evaluated over some finite period of time. In some cases, the energy consumption can be monitored as changes in the power level corresponding to changes in the instantaneous energy consumption. The measurement procedures and conditions including the exact location of these measurements are often standardized, or they are specified by the vendors or operators [12]. For instance, different energy metrics are recommended for wired and wireless networks, respectively, and the metrics design is also influenced by the network architecture, workload or utilization, and the services offered. In wireless systems, the energy consumption is strongly affected by the physical layer techniques such as multiple antennas, modulation and channel coding used [13]. In radio access networks, the base stations are the largest energy consumers. The base station architecture including the power amplifier design and antenna mounting, exploiting cooperative communication techniques within the cell using relays, optimizing the cross-layer design, and distributing traffic load with service differentiation are some popular strategies to reduce the operational energy demands of mobile networks [13].

The most commonly used energy and power metrics for green mobile networks are summarized in [4]. The perspectives of network operators and mobile users are considered to ensure the energy consumption fairness between these two stakeholders. The energy efficiency of radio equipment is defined and treated separately from the energy consumption of the whole network. The energy efficiency trade-offs against the spectral, deployment, routing, scheduling, packet length, and delay efficiency are surveyed in [14]. A comprehensive survey of energy metrics for the radio access networks is provided in [12]. The energy metrics concerned with the operational energy consumption are summarized in Table III. Even though the embodied energy can be significant and often exceeds the operational energy [15], the embodied energy has not been consider in [12], since it is estimated from the physical models rather than measured. The Wireless Sensor Networks (WSNs) are used as an enabler in many applications despite their energy conservation issues, and the requirement for long time deployment and a compact size. This problem can be mitigated by optimizing the communication protocols [16], power control, and other methods [17]. The energy consumption is also critical to maintain a topology lifetime of Mobile Ad-hoc NETworks (MANETs) [18]. The energy consumption calculations for the transceivers in wireless networks are outlined in Table IV [19]. The mobile phone users usually have little knowledge of how the battery charge is used [20], and how it is affected by the software and hardware components [21,22]. The maximum output power for mobile phones is set by the regulators to 3 Watts [23]. On the other hand, the evaluation of energy consumption in wired networks which are rarely reliant on the batteries has received much less attention in the literature [24]. For instance, the energy-aware routing and management for wired networks has been considered in [25].

Table III
ENERGY AND POWER METRICS FOR RADIO ACCESS AND MOBILE NETWORKS

| Key factors | |
|---|---|
| • network load (typically 0, 50 and 100% traffic)<br>• performance constraints (e.g. QoS)<br>• network topology and infrastructure | • segment to consider (node, equipment, link)<br>• protocols and applications<br>• mode (transmission, reception, standby, sleep) |
| Measurements | |
| • relative to reference system or technique<br>• absolute energy (expressed in Joules) | • time period of averaging<br>• measurement port location |
| Total energy consumed | |
| $E_{total} = E_{operational} + E_{embodied}$ [Joules]<br><br>$E_{operational} = (P_{RF} + P_{overhead}) \times T_{run\text{-}time}$ | • assume $P_{RF}$ proportional to rate<br>• assume $P_{RF}$ and $P_{overhead}$ are independent |
| Embodied energy | |
| $E_{embodied} = E_{manufacturing} + E_{transport} + E_{installation}$<br>$\qquad + E_{maintenance} + E_{disposal}$<br>$E_{maintenance} / year \approx 1\%$ of $E_{manufacturing}$ | • life-cycle assessment (LCA)<br>• environmental KPI (KEPI)<br>• ecological footprint analysis (EFA) |
| Operational energy | |
| Energy consumption rating (ECR)<br><br>$\qquad ECR = P_f / T_f$ **[Watts/bps]**<br><br>Energy efficiency rate (EER)<br><br>$\qquad EER = 1 / ECR$ [bps/Watts] | • $P_f$ is peak power and $T_f$ is maximum throughput<br>• this metric is not standardized yet |

Variable load ECR (ECR-VL)

$$ECR_{VL} = \frac{\alpha \cdot P_{100} + \beta \cdot P_{50} + \gamma \cdot P_{30} + \delta \cdot P_{10} + \epsilon \cdot PI}{\alpha \cdot T_{f} + \beta \cdot T_{50} + \gamma \cdot T_{30} + \delta \cdot T_{10}}$$

- account for dynamic network power management
- $P_{xx}$ is equipment power under load, $PI$ is idle power, and $T_{xx}$ are corresponding throughputs
- $\alpha, \beta, \gamma, \delta, \epsilon$ are weighting factors

ECR for radio access networks (RAN)

- $E_{cell}$ is energy expended over time T
- $M_{cell}$ is # application bits transported over time T
- $P_{cell}$ is the cell power, and $S_{cell}$ is transmission rate

| ATIS network interface |
|---|

Transport equipment

$$\frac{-\log(P_{total})}{Throughput} \ [dB/Gbps]$$

Switches and routers

$$\frac{\log \ P_{total}}{Forwarding \ Capacity} \ [dB/Gbps]$$

Access equipment

$$\frac{\# \ Access \ Lines}{P_{total}} \ [Lines/watt]$$

Power ratio

$$\frac{P_{out,total}}{P_{in,total}} \ [no \ unit]$$

Power amplifier

$$\frac{P_{RF,out}}{P_{total,in}} \ [no \ unit]$$

Radio base station

$$P_{Site} = PC + PRRH \ \ [Watts]$$

| ITU metrics |
|---|

Wired networks

$$\frac{Power}{Subscriber \times Traffic \times Distance} \ [Watts/bps/m]$$

Normalized energy per transmitted bit

$$\frac{P \times T}{D} \ [Joules/bit]$$

- transmission of D bits in T seconds at power P

Wireless networks

$$\frac{Power}{Subscriber \times Traffic \times Area} \ [Watts/bps/m^2]$$

Power per line for broadband equipment

$$Power/line = \frac{P_{line}}{N_{subscribers}} \ [Watts]$$

| Cellular systems |
|---|

$$KPI_{Rural} = \frac{A_{Coverage}}{P_{Site}} \ \ [km^2/Watts]$$

$$KPI_{Urban} = \frac{N_{BusyHour}}{P_{site}} \ \ [subscribers/Watts]$$

Table IV

POWER AND ENERGY FOR THE TRANSCEIVERS IN WIRELESS NETWORKS

| Transmission (T) and reception (R) mode | energy to transmit-receive one data packet [Joules] $$E_{T-R} = \frac{current \ drain \ [A] \times power \ supply \ [V] \times \#bits}{bandwidth \ [Hz]} = P_{T-R} \cdot T_{T-R}$$ |
|---|---|
| Idle (I) and overhearing (O) mode | power for constant listening to wireless medium to detect next packet $$P_I = P_O = P_R$$ |

More recently, the energy consumption monitoring has become important also in computing applications [26]. A comprehensive view on the energy consumption in terms of the greenhouse emissions in ICT systems is provided in [27], and the energy savings due to the use of ICT in transportation, material production and manufacturing has been considered in [28].

## 3.    QUALITY OF SERVICE (QOS) METRICS

The ITU (International Telecommunication Union) and ETSI (European Telecommunications Standards Institute) **define the QoS as** "t**h**e ability of a network or network portion to provide the **functions related to communications between users**" [29]. The QoS is defined in [30] **as "totality of** characteristics of a telecommunication service that bears on its ability to satisfy stated and implied needs of the user of the service." In general, telecommunication networks maintain the desired level of QoS by monitoring different KPIs. However, it is possible to change or update the KPIs and target the same QoS. The Internet Service Providers (ISPs) do not explicitly sell the QoS to their subscribers and business clients, but the QoS is included and offered at different costs in their subscription packages [31]. The regulatory bodies set forth the policies to ensure fair pricing for the services provided by the ISPs to satisfy the subscribers expectations. These policies must be agnostic of the network architecture, and they must account for different traffic classes such as real-time and non-real-time services which must be treated differently [32]. For instance, non-real time services may be concerned with the reliable packet delivery. The end-users normally expect the QoS performance for applications over both wired and wireless connections [33]. Guaranteeing the QoS is, generally, challenging due to dynamic channel allocations, and the energy savings and fault tolerance mechanisms used. Therefore, different QoS classes can be considered to better match the network architecture and the used protocols to the application requirements. For example, QoS classes can be supported by a cross-layer design in WIMAX networks [33], by optimizing the routing decisions in ad-hoc networks [34], and by call admission control (CAC) and mobility prediction in the cellular networks [35]. The distributed channel allocation may help to guarantee the end-to-end QoS. The design of KPIs for QoS in the 5G networks is investigated in [36] where it is suggested that the QoS can be achieved by focusing on network function virtualization (NFV). The QoS for mobile users may incorporate different metrics such as a source-destination route stability [37].

More importantly, the current architecture of the Internet has been designed to offer best-effort services without providing any assurances of their quality. Nevertheless, passive or active traffic monitoring can used to determine the provided QoS [38,39]. In some occasions, the network may offer priority for certain traffic flows to adjust their QoS while not ignoring lower priority traffic. Prioritizing traffic, and more generally, traffic engineering is achieved by reserving and appropriately managing the network resources, usually at the access routers and switches. In IP-based wired networks, IntServ and DiffServ frameworks are used for provisioning of the QoS [40]. In particular, the integrated service (IntServ) is a fine-grained end-to-end QoS prioritization mechanism for individual uni-directional traffic streams [41]. A recent demand for the high-speed routes at the network core which has to support large number of concurrent traffic flows favors a differentiated services (DiffServ) approach [40,42]. This traffic framework uses a coarse based traffic classification and prioritization between the routes at the core [43]. The QoS may be also provided for aggregated rather than individual traffic flows [44]. In order to account for possible network congestion and a mixture of different service types, DiffServ can be combined with other traffic engineering mechanisms [45]. The QoS service classes supported by IntServ and DiffServ are summarized in Table V [46]. More importantly, note that the standardization bodies do not consider QoS requirements [33] in order to support the integration of heterogeneous network architectures.

Table V
QoS PROVISIONING FRAMEWORKS IN IP-BASED WIRED NETWORKS

| IntServ | | DiffServ | |
|---|---|---|---|
| Elastic: | no quality guarantees ("best effort") | Premium: | low delay and losses, guaranteed bandwidth |
| Tolerant real-time: | delay sensitive and high bandwidth | Assured: | like Premium but relaxed requirements to delays and losses |
| Intolerant real-time: | low delay and guaranteed bandwidth | Olympic: | no delay requirements |

The QoS metrics in both wired and wireless networks can be classified as application-based QoS (AQoS) and network-based QoS (NQoS). The former is concerned with the end-to-end quality of real-time applications such as voice and video. The latter considers the traffic quality enabled by the network equipment such as the routers and switches. In other words, AQoS is a measure of the user satisfaction while NQoS relates to the network capability to deliver a service. Most QoS measurements correspond to the NQoS metrics [47]. The overall QoS is simply a sum of AQoS and NQoS.

## A. AQoS METRICS

The QoS at the application layer is driven by the human perception which is based on three main characteristics: the spatial and temporal perceptions for video, and the acoustic frequency range for voice.

Thus, AQoS metrics reflect the QoS for the end-users such as a content resolution and the frame rate. The QoS requests from applications can vary over time, but all these requests should be recognized and well-defined. However, the application requirements of users are usually expressed using a non-technical language which is then translated into corresponding network parameters. The resulting network parameters may be very restrictive or more tolerant. The related classification of applications is given in Table VI [48,49]. In [50], the AQoS metrics and the underlying parameters for each application are defined, and then used to enhance the user experience as well as to improve the utilization of network resources.

Table VI
APPLICATION CLASSIFICATION

| Elastic, non-interactive: | file download, email | ITU-T classification: |
|---|---|---|
| Elastic, interactive: | web browsing | • audio |
| Non-elastic, non-interactive: | Tv and VoD, streaming, broadcast | • video |
| Non-elastic, interactive: | VoIP, video conference, online gaming | • data related |

## B. NQoS METRICS

The NQoS metrics are directly concerned with the services provided by the network and networking equipment in order to improve the end-to-end QoS. The quantifiable metrics to define the QoS performance of the network services are: throughput, delay, packet loss and availability as reviewed next.

The throughput (in bits/s) is the average number of successfully delivered bits per time slot over a communication channel [51]. The key factors affecting the throughput are the number of hops, node mobility and transmission range [52]. In wireless systems, the throughput is constrained by signal-to-noise ratio (SNR) at the receiver which can be optimized by the choice of transmission signaling including modulation, coding, and packet size [53,54]. At the network level, the appropriate routing strategy for packet aggregation and forwarding [51], a network coordination function [55], or using optical backhaul to reduce the number of hops [56] can increase the end-to-end throughput. In wireless networks, suppressing the interference by directional antennas [57] and optimizing the communication protocols [58] will enhance the overall network throughput. The throughput metrics from the literature which are used in simulations as well as for direct measurements are summarized in Table VII. In computer science, the computing throughput is a quantity of work that a computer can deliver within a certain time period [59].

<div align="center">

Table VII

THE THROUGHPUT METRICS

</div>

| Transmission time | | Throughput | |
|---|---|---|---|
| $\dfrac{\text{File size}}{\text{Bandwidth}}$ [sec] | | $\dfrac{\text{File size}}{\text{Transmission time}}$ [bps] | |
| Link throughput | | Packet Delivery Ratio (PDR) | |
| $\dfrac{\sum \#\text{bits}}{T_{total}}$ [bps/link] | | $\dfrac{\sum N_{received}}{\sum N_{generated}} \times 100$ [%] | |
| Average end-to-end delay (AED) | | Total network throughput [60] | |
| $\sum\limits_{packets} T_{to\ receive} - \sum\limits_{packets} T_{to\ send}$ [msec] | | $\dfrac{\text{Segment size}}{T_{round\text{-}trip} \times P_{frame\ error}}$ [bps] | |

The throughput in the IP-based networks is influenced significantly by the distance, network architecture and topology, and the network load. The throughput between the end-user and the service provider is a key parameter in the SLAs which, among other things, often specify the mean time to failure (MTTF), to repair (MTTR) and between failures (MTBF), respectively [61]. In order to deliver the agreed SLA, the network performance measurements and the network resources control policy are defined to deliver the required QoS [62]. The dependencies of the SLAs among the different parties are shown in Fig. 3.

The delay, as defined by Cisco [63], is "the finite amount of time it takes a packet to reach the receiving endpoint after being transmitted from the sending endpoint. " The delay usually dramatically increases with network congestion, and it negatively affects the perceived QoS. The propagation delay can normally be neglected. However, the buffers at routers first store the whole packet before this packet is forwarded further which doubles the corresponding delay for each intermediate hop (see Table VIII). The average packet delay is referred to as a latency, and it is usually one of the main QoS performance metrics included in the SLAs.

Table VIII
POINT-TO-POINT AND END-TO-END DELAYS

| | |
|---|---|
| $$D_{nodal} = D_{proc} + D_{queue} + D_{trans} + D_{prop} \ [sec]$$ | |
| $D_{proc}$ | processing delay (error checking, reading header, routing decision) |
| $D_{queue}$ | queuing delay until transmission ($L_{queue}$ is average length of the queue) |
| $D_{trans}$ | $$D_{queue} = D_{trans} \times L_{queue}$$ transmission delay until last packet bit sent out (L is packet length, R is rate) $$D_{trans} = L \ / \ R$$ |
| $D_{prop}$ | propagation delay for the bit to reach destination (D is distance, S is speed) $$D_{prop} = D \ / \ S$$ |
| | where speed is 186,000 miles/sec, and for copper & fibre 100,000 miles/sec |
| other | link capacities, traffic load, keeping track of route changes, clock synchronizations |

The packet delay can be measured point-to-point in one direction only [64]. The acceptable values for such delays can range from milliseconds for Real-Time Interactive (RTI) multimedia applications to minutes for Video-on-Demand (VoD) applications [65]. The delay can be also defined as a round-trip time (RTT), or end-to-end delay between the user end-device and the service provisioning point to **measure the time between a request and the response. The "ping" command is a popular tool for** measuring these delays, and possibly to optimize the routing. Comprehensive delay measurements including their variations, path symmetry and queuing latencies are reported in [66] and [67].

**The delay variations or delay jitter are "the variations in latency for packets within** a given data **stream"** [68]. The jitter notion and its causes are more complex in wireless than in wired networks. For instance, except normal network congestion, the wireless networks suffer from the interference and time-varying propagation conditions. If the difference between the maximum and minimum RTT values exceeds a certain threshold, the real-time applications may experience a significant degradation of quality or even become unusable. The jitter for RTT values can be measured as a standard deviation, i.e., the square-root of sample variance. The average jitter at the router can be defined for receiving the packets as well as for transmitting the packets. In IP-based networks, the jitter can be reduced by congestion control mechanisms [69], by removing the bottleneck links (e.g., those emanating from the gateway router), and by buffering data packets to filter out the packet arrival variations [70]. The network resources reservation can be also used to effectively suppress the jitter [71]. The jitter classification is summarized in Table IX [72,73].

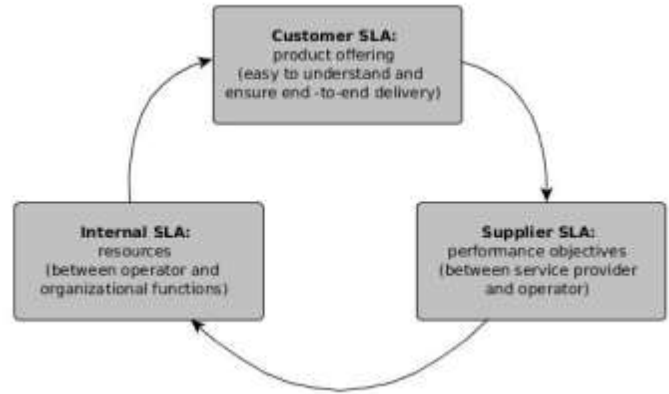

Fig. 3: The SLA party dependencies [74].

Table IX

JITTER CLASSIFICATION

| | |
|---|---|
| Random jitter: | referred as background or thermal Gaussian noise |
| Deterministic jitter: | periodic jitter or data dependent jitter |
| Correlated jitter: | usually periodic jitter (e.g. serial data interfere with clock signal) |
| Uncorrelated jitter: | no cause or influence, typical for random jitter |
| Constant jitter: | packet to packet delay variations |
| Transient jitter: | in connection with substantial incremental delay |
| Short term delay variations: | commonly associated with congestions and route changes |

The packet loss, "the failure of one or more transmitted packets to arrive at their destination " [75], is another important QoS performance metric. It is especially meaningful for real-time interactive applications requiring high data throughput with low latency [76]. The packet losses result from the buffer overflows during network congestion, from the link and equipment failures, and from the transmission errors due to signal propagation effects over an unreliable medium and the electromagnetic inference. The packet loss rate crucially depends on the packet size and the transport protocol used [77]. The widely used transport protocols, TCP and UDP, in IP-based networks are compared in Table X. Assuming a single link or one hop transmission, the packet loss metrics can be defined similarly to the throughput metrics, however, measuring the end-to-end packet losses (e.g., using active network probing) is not straightforward as indicated in [78-81]. Simple Network Management Protocol (SNMP) and the "ping" command (which exploits ICMP echo messages) can be used to passively monitor the network routers and infer the packet losses [82]. There exist also more sophisticated network probing and network sampling methods such as Poisson-modulated probes, unicast probes, ZING tool, and PASTA principle which are described in [82], [83] and [84], respectively. The accuracy of these methods depends on the characteristics and interpretation of the collected data.

Table X
COMPARISON OF THE TCP AND UDP TRANSPORT PROTOCOLS

| TCP protocol | UDP protocol |
|---|---|
| <ul><li>detect packet losses</li><li>error correction (but redundancy)</li><li>byte-oriented flow</li><li>reliable in-order packet delivery</li><li>delay and jitter  (retransmissions)</li><li>congestion control</li><li>difficult for real-time services</li><li>idle session timeout: 60 min</li><li>connection open/close cost: 300 bytes</li><li>packet header overhead: 20 bytes</li><li>usually used with HTTP, FTP, SMTP</li></ul> | <ul><li>no detection of packet loss</li><li>no guarantee, possibly out-of-order delivery</li><li>minimize delay and jitter</li><li>no congestion control</li><li>usable for real-time services</li><li>Idle session timeout: 5 min</li><li>connection cost: 28 bytes</li><li>packet header overhead: 8 bytes</li><li>fast data transmitting (streaming)</li><li>usually used with DHCP, DNS, TFTP, SNMP, RIP, VoIP</li></ul> |
| Both: packet-based, connectionless, best-effort | |

The availability is the last QoS metric we consider in this subsection. According to ISO 2382-14, it is "the ability of a functional unit to be in a state to perform a required function under given conditions at a given instant of time or over a given time interval, assuming that the required external resources are provided " [85]. Generally, despite failures of the networking equipment, the whole network needs to remain reliable and assure continuity of the service with acceptable level of performance. In practice, the network operators typically aim to achieve the availability as high as 99.999% ("five-nines") of the service uptime [86]. However, there is great uncertainty in knowing the reliability of every network component including the hardware, software and the external power supply, so measuring these reliabilities is not straightforward. However, the availability upper bounds for the whole system can still be obtained [87].  The SLA normally specifies the availability levels which can include the acceptable downtime and outage periods during a specified time interval.

From the user perspective, the point-to-point availability is important and it can be measured directly whereas, for the network management, we must consider the network-level availability. Many metrics have been defined to quantify the availability of different network elements as shown in Table XI [88-90]. The fundamental difference between the availability of wired and wireless networks is related to the dynamics of these two types of networks [91]. For instance, wireless networks are more vulnerable due to detrimental propagation conditions and mobility of nodes, however, they are usually also cheaper to deploy and maintain. On the other hand, in wired networks, the software and hardware problems are main sources of failures. The network availability is sensitive to network topology which is characterized by the average shortest path, network diameter and the number of shared links [92]. A hybrid networks combining optical and wireless links can be a cost-effective solution offering also a good throughput and availability [93]. Another option is to use overlay wireless networks to off-load traffic in case of the optical link failures [94].

Table XI
AVAILABILITY METRICS

| | |
|---|---|
| Mean time to failure (MTTF): | $MTTF = E[t] = \int_0^\infty t\,f(t)\,dt \approx \dfrac{\sum_i T_{i,downtime} - T_{i,uptime}}{\# \text{ breakdowns}}$ [hours] |
| Mean time to repair (MTTR): | $TTR = \dfrac{\text{Total down time}}{\# \text{ failures}}$ [hours] |
| Impacted user minutes (IUM): | $IUM = \# \text{ affected users} \times T_{incident}$ [min] |
| Defects per million (DPM): | $DPM = \dfrac{\text{Total user time-Total user outages time}}{\text{Total user time}} \times 100\%$ |
| Mean time between failures (MTBF): | $MTBF = \dfrac{\sum T_{uptime}}{\# \text{ breakdowns}}$ [hours] |
| Point availability: | $A(t) = \int_0^t R(t-u)m(u)\,du$<br>• last repair at time u, m(u) is renewal density of the system<br>• probability the system is operational at random time t>u |
| Average uptime availability: | $\overline{A(t)} = \dfrac{1}{t}\int_0^t A(u)\,du$<br>• proportion of time the system is available for use |
| Steady state availability: | $A(\infty) = \lim_{t\to\infty} A(t) = \dfrac{\mu}{\lambda+\pi} + \dfrac{\lambda}{\lambda+\pi}$<br>• availability limit as time goes to infinity<br>• repair rate $\mu$ and failure rate $\lambda$ are assumed constant |
| Inherent availability: | $A_i = \dfrac{MTTF}{MTTF + MTTR}$ , or $A_i = \dfrac{MTBF}{MTBF + MTTR}$ (for systems)<br>• steady state availability considering only corrective downtime |
| Achieved availability: | $A_A = \dfrac{MTBF}{MTBF + \overline{M}}$<br>• availability with planned shutdowns<br>• $\overline{M}$ is mean maintenance downtime |
| Operational availability: | $A_O = \dfrac{T_{uptime}}{T_{operation\ cycle}}$ |

- average availability with all expected sources of downtime

The network or service availability must often trade-off other important metrics, especially in more complex systems such as the emerging 5G networks and cloud computing [95]. For example, the wireless networks for e-health applications have to balance the availability and accessibility against the security and privacy [96]. In equipment monitoring applications such as in oil and gas industry, the network availability has direct effect on the economic losses, environmental damage and safety [97]. Particularly in wireless networks, the network availability can be improved by adding redundant nodes as shown in [98]. The required level of redundancy to achieve a given availability in WSNs is determined in [99].

## 4. QUALITY OF EXPERIENCE (QOE) METRICS

The ITU defines the QoE as "the overall acceptability of an application or service, as perceived subjectively by the end-user" [100]. The QoE is used to measure and express, preferably as numerical values, the experience and perception of the users with a service. However, the QoE is more than just an indicator as it explores the users' level of satisfaction and their response to the service. The ambiguity of the QoE metrics stems from a lack of realistic models of the human behavior. It is an important criterion when considering how to design the whole ecosystem of telecommunication services as indicated in Fig. 4. More importantly, the metrics in this ecosystem have to continuously evolve in order to keep the pace with the development of the underlying ICT [101]. The QoE metrics can be associated with both subjective and objective quality needs, and they go beyond rather than just complement the technical performance indicators defined by the QoS metrics. For example, as the 5G systems are transforming from the network-centric to user- or human-centric architectures, the network management shifts from the QoS to the QoE oriented performance evaluations [102]. However, the QoE is not entirely independent from the QoS metrics. The network operators can use the QoE metrics to understand how to improve their services and set the adequate pricing levels to optimize their economic returns as most users prefer affordable services that are priced fairly. A customer survey revealed that almost 90% of the users would switch the service provider rather than to complain [103]. This was evidenced by Vodafone in Australia when they lost nearly half a million subscribers to other operators in 2011 [104].

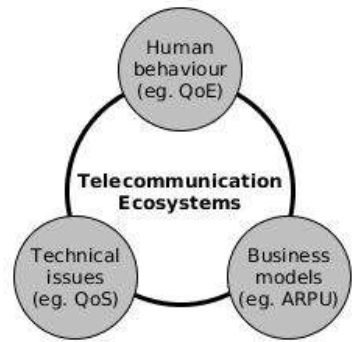

Fig. 4: The ecosystem of telecommunication services.

The users' perception of a service is altered when they notice a change in some service characteristics which is usually a result of some technical upgrade [105]. In some cases, such upgrade may actually deteriorate the QoE. It is therefore important to evolve the network, so that both the QoS and the QoE are improved at the same time. A separate study showed that the QoE should consider also other factors such as aesthetics, visual appeal, and friendliness, organization and usability of the user interface [106]. All these factors affect the end user assessment of the service, however, their influences are not the same. For example, the content type has an impact on the perceived service performance, but it does not affect the perceived aesthetics. Many factors affecting the perceived quality of the adaptive video streaming are categorized in [107]. The authors consider the subjective QoE metrics assuming both the human-computer interactions and the network-centric frameworks, and also explore how to map technical aspects of the video streaming onto the anticipated QoE. Since the multimedia content is directly determined by the underlying data encoding and processing, it has been suggested in [108] to define the numerical QoE metrics as a function of the AQoS and NQoS values. The QoS-aware mobility and routing mechanisms can alleviate packet losses which enhances the user perception measured as the Mean Opinion Score (MOS) metric [109]. In practice, the relationship between the QoS and the QoE is more complicated, and there is no guarantee that improving the QoS would necessarily result in a better QoE. For example, enhancing the QoE by improving the throughput, delay and jitter may be unsuccessful [99]. The relationship between the QoS and the QoE for subjective as well as objective assessment of the video quality is investigated in [110]. Different approaches to QoE and particularly a data-driven QoE modeling is analyzed and compared. The objective and subjective speech quality assessment frameworks are surveyed in [111]. The most important factors affecting the speech quality are considered across all layers of the protocol stack. The main objective is to identify suitable metrics for automated evaluation of the QoE for VoIP systems in mobile networks. A comprehensive survey of the QoE measurement methodologies for online and on-demand video streaming services has been carried out in [112]. In study [113], it is shown that high-quality video streaming may actually yield poor QoE due to frequent re-buffering during the playback. The affected users are more likely to complain than accept a lower-quality video. Consequently, the network operators and content providers should cooperate and monitor not only the network performance related metrics in order to provide a good QoE for the end-users, and to reduce their complaints. For example, the YouTube video streaming automatically adjusts the video quality to the connection speed. The video streaming is also an attractive option in marketing the subscription packages by the network operators. The user perception and requirements are driven by other factors such as their individual needs and behaviors, the user interactions (among themselves and with the network), the service provisioning efficiency and accuracy, and the context [101,114,115]. The context can be defined as "any information that can be used to characterize the situation of an entity (e.g. a user, a network or a device)" [116]. This definition allows for both the static and the dynamic contexts. Therefore, finding the comprehensive solutions to offer the satisfactory QoE in a wide range of scenarios is, generally, a very challenging task. It is unclear how to identify, interpret and measure the context, and how to accurately emulate its dynamics for the application development and testing purposes and within the network testbeds [117]. For instance, the QoS parameters vary over the day and with the user location.

In addition to the QoE, we may consider the related user experience (UX) metrics defined as "a person's perceptions and responses that result from the use or anticipated use of a product, system or service" [118]. In spite of the UX being recommended by the regulatory bodies such as ITU and ISO, the authors in [119] pointed out the difficulty of the UX with its dynamic and subjective nature. The UX is affected by the user equipment (UE) capabilities such as the screen size and resolution. In a user-centric design, the UX should jointly consider the experiences with the service, product and the brand.

The situation is even more complicated in the heterogeneous networks where the QoE is influenced by the content type, service type, pricing policy and the psychological characteristics [120]. The end-to-end QoE guarantees require a common platform in spite of a diverse range of services offered by these networks over various media and employing various networking solutions. This drives the need for the continuous development, technology adoption and new business and service models. The average revenue per user (ARPU) and the customer churn are the main metrics that tie the customers with the service providers. The current research is focusing on obtaining good QoE models and on the QoE forecasting [104].

In practice, the QoE is assessed using either the subjective or objective metrics. We will review these metrics next.

A.  SUBJECTIVE ASSESSMENT OF QoE

The subjective QoE metrics take the human perspective [104,121]. They use a subject test to collect data and evaluate the user opinions. They can be further subdivided into quantitative and qualitative metrics where the former usually represents the directly measured statistics while the latter assesses the human perception through surveys and interviews. It has long been known that the human perception is exceptionally precise at grasping the differences rather than sensing the direct values [122]. The subjective QoE metrics are frequently used in practice, however, the caveat is that their evaluation is time consuming and laborious, and they cannot be used for real-time service monitoring [123-125]. Neither the standardization bodies such as the ITU-T and the ETSI consider the subjective QoE metrics and measurements for new, real-time applications [126]. However, both the content providers and the network operators are now focusing on the QoE subjective metrics in order to offer meaningful multimedia experiences and to overcome limitations of the more traditional QoS-aware network designs. This approach is beneficial to accurately forecast the customer satisfactions, even though it is still very complex, especially in real-time scenarios. A survey of the QoE subjective metrics is provided in Table XII.

Table XII
QoE SUBJECTIVE METRICS

| Subjective metric | Features | Application |
| --- | --- | --- |
| Mean Opinion Score (MOS) [ITU-T P.800] | numerical QoE using subjective tests (Absolute category rating scale), but ignores other important aspects (e.g. degree of interactivity) | video and audio |
| Double Stimulus Continuous Quality Scale (DSCQS) [ITU-R BT.500-11] [127] | index of video quality which is less sensitive to context, but also inefficient for real-time evaluations | pictures, TV, video stream, multimedia |
| Single Stimulus Continuous Quality Evaluation (SSCQE) [127] | more representative quality estimates for quality monitoring applications | TV, video stream, and multimedia |
| Absolute Category (ACR), optionally with Hidden Ref. Removal (HRR) [128] | efficient, reliable and standardized method permitting great number of test conditions in a single test period | video streaming |
| Double Stimulus Impairment Scale (DSIS) [ITU-R BT.500-11] | paired evaluation: unimpaired reference video against impaired video | pictures and video |
| Single Stimulus Continuous Quality Evaluation (SSCQE) [ITU-R BT.500-11] | use of a slider device with no standard video | video |
| Single Stimulus (SS) | also referred as Absolute Category Rating (ACR) | video |

| Just Noticeable Difference (JND) scale [122] | series of comparison tests on two samples while intensity in one sample increases or decreases | video |
|---|---|---|
| Maximum Likelihood Differ. Scaling (MLDS) [129] | relative difference in quality to represent the utility of the tested parameter on visual quality | images and video |

## B. OBJECTIVE ASSESSMENT OF QoE

Unlike the subjective QoE metrics that directly evaluate the human perception, the objective QoE metrics utilize data, algorithms, and models to infer the user satisfactions. The data may be provided by applications or by the network protocol layers including the AQoS and NQoS measurements [130]. The objective modeling of system quality is attractive for its low implementation requirements, adaptivity, and ability to operate in real-time settings, and it is used extensively by the network operators, codec engineers and the application developers.

The objective QoE metrics can be divided into three classes depending on the extent of use of the original reference signal such as image or video [131]. The full-reference (FR) metrics calculate deprivation of the image at the decoder output by making pixel-to-pixel comparison with the same image at the encoder input. The reduced-reference (RR) metrics make the comparison for a part of the image. Since the original undistorted image or its part may not be available or is impractical to obtain, the no-reference (NR) metrics mimic the ability of the human observers to determine the quality of the observed image. The challenge in devising the specific QoE objective metrics is often unknown dependence on the system parameters, non-linear nature of the human perception, and the varying satisfaction of the users over time. For instance, the users may become more acquainted with the service after some initial time period, or they become less satisfied [132], so the QoE metrics must be monitored and evaluated periodically.

One of the most commonly used QoE objective metrics is peak signal-to-noise ratio (PSNR) [133]. It can be used to determine the quality of video streaming services and to expose their impairments as these services continue to dominate traffic in the future broadband networks [36]. The PSNR is effectively an inverse value of the mean square error (MSE), and it is also the simplest FR type of the QoE metric. However, since the PSNR metric has been shown to be inconsistent with the human visual perception, more complex FR metrics were developed such as structural similarity index (SSIM) and perceptual evaluation of video quality (PEVQ). The latter metric is part of a family of metrics that are now standardized by the ITU-T for the quality assessment of various multimedia services. Similarly to PSNR, the SSIM metric performs frame-to-frame comparisons between the original and the displayed video, but the comparisons are based on the luminance, contrast and structure of each frame. The SSIM outputs decimal values between 0 and 1. If the frame after the decoding matches the source frame, there is no distortion and the index value is 1. Likewise, small SSIM values indicate significant distortion of the video content. The video quality metric (VQM) is another FR metric system. It yields the integer values between 0 and 5 with 5 being the best quality. This metric tries to reflect the subjective human perception by considering the image noise, color blurring and pixel deformations. In some scenarios, particularly during the development of objective service quality models, the objective metrics such as PSNR and SSIM are mapped to subjective scores, usually assuming MOS as indicated in Table XIII [133].

TABLE XIII

MAPPING OF PSNR LEVELS TO MOS SCORES

| PSNR [dB] | MOS | |
|-----------|-----|-------------|
| > 37 | 5 | (excellent) |
| 31 - 37 | 4 | (good) |
| 25 - 31 | 3 | (fair) |
| 20 - 25 | 2 | (poor) |
| < 20 | 1 | (bad) |

For interactive real-time applications, numerous studies proposed to replace the subjective and objective video quality metrics with the QoE metrics that directly reflect the current network conditions. For instance, in a video Skype call, the users maintain the communication even if the video quality temporarily drops. Inferring the current perceived QoE is important to adjust the video streaming parameters such as the resolution and a frame rate. The objective metrics for video quality assessment are surveyed in Table XIV.

Table XIV

QoE OBJECTIVE METRICS

| Objective metric | Features | Application |
|------------------|----------|-------------|
| E-Model [134] | estimates MOS in real-time (computes the R-factor) | audio (ITU G.107), VoIP |
| Perception Evaluation of Speech Quality (PESQ) model [135] | estimates MOS by comparing reference and received signals | audio-visual communications, FR method |
| Application Performance Index (APDEX) [136] | consumer satisfaction on scale between 0 (no users satisfied) to 1 (all users satisfied) | any performance evaluation for end-users |
| MOVIS project [137] | a method to estimate QoE for single and | online measurements of end-user service quality for |

| | group consumers | streamed content |
|---|---|---|
| Peak Signal-to-Noise Ratio (PSNR) [65] | similarity between two different images defined using Mean Squared Error (MSE) $$PSNR = 10 \log_{10}\left(\frac{max(image)^2}{MSE}\right)^2$$ | simplest objective video quality, however, no consideration of human perception |
| Moving Picture Quality Metric (MPQM) | video quality using a mix of content dependent factors as well as network impairments (e.g. packet losses, delays) | video, IPTV, NR and RR methods, real time quality monitoring, but less accurate |
| Motion-based Video Integrity Evaluation (MOVIE) [138] | impairments considered jointly in space and time | FR method, but high computational and memory costs |
| Structural Similarity Index (SSIM) [139] | degradation of structural info such as luminance and contrast | images, video, FR method |
| Video Quality Metric (VQM) [ITU-T G.1070] [65] | detects human perceivable artifacts in images and video based on codec type, block and color distortions etc. | video, FR method |
| Pseudo Subjective Quality Assessment (PSQA) [140] | real-time evaluation of quality of communications over packet networks | video and audio |
| Bayesian networks and context-aware state-space approach [141] | sequential modeling to monitor QoE over time | multimedia |
| User Satisfaction Index (USI) [142] | rigorous analysis of source and network level QoS metrics over call duration | interactive real-time multimedia |

To summarize our discussion about the QoE metrics, we point out some of the main challenges. The evaluation of the subjective QoE metrics is costly and cannot be used with real-time applications such as popular video streaming services. The objective QoE metrics do not exactly reflect the sentiment of users, and they can only exploit the measurements that are available. In order to improve the predictive accuracy of the objective QoE metrics, a hybrid approach was suggested in [124], [143] and [144]. The reference [145] suggests collecting all measurements from the whole system, evaluating the QoS of the network, and then mapping it to the QoE. However, some subjective parameters such as the user mood are usually completely unrelated to the state of the network. Although some methods for inferring the user mood are being considered, the main strategy prevailing today is to match the user expectations by designing networks that are more flexible and better utilize their resources [146].

# 5.   SECURITY METRICS

The ISO definition states that "network security is the security of devices, and security of management activities related to the devices, applications, services, and the end-users, in addition to security of information being transferred across the communication links " [147]. The security of ICT services has now influence even on the world geopolitical landscape [148]. Unlike other metrics discussed so far, the metrics related to security do not impact the user immediate satisfaction with the service, so the users are unaware of the actual security level. The service providers are also reluctant to disclose any security issues in order to retain their subscribers. This creates a major drive for defining the independent security metrics that can be used without interference from the providers and the authorities.

In general, the security metrics aim to assess the vulnerability of the end-users to any unauthorized access, theft of sensitive information, and privacy invasion. The objective is to guarantee business continuity and to prevent or minimize the damages caused by the security incidents. The security dimensions span technical, organizational, human and any other aspects that may affect the offered services and products. The security measures can be evaluated economically either as a return on investment (ROI) to measure the efficiency of the security improvements as a result of the business investments, or as a reduction of the financial losses due to the security breaches [149]. The security metrics are also needed to make preemptive decisions about protecting the systems, resources, and data. In the United States, the government agencies are required to integrate security metrics into all their programs [150]. The implementation strategies and analysis of the security metrics are investigated in references [151-153]. However, in many scenarios, it may be difficult to identify the appropriate security metrics even though we may try to enumerate all major threats and vulnerabilities. For example, the major threats and vulnerabilities for cognitive radio networks have been identified in [154] and for software defined networks in [155], but none of these papers derive or provide any specific security metrics. The probabilities of different attacks in information-centric networks are given in [156]. The effects of security attacks may be quantified as having low, medium or high impact whereas most other papers on security only consider whether the attack or threat is or is not present or ongoing. However, no quantitative security metrics are provided nor derived which may again indicate that such metrics are not easy to define, measure and validate.

The standardization bodies (e.g., ITU, ISO/IEC, 3GPP/2, and ETSI) currently do not specify any specific security protection or intrusion-detection methods, so such methods have to be devised and implemented by the network operators in collaboration with the network infrastructure providers. On the other hand, the specialized independent security bodies and institutions often define the levels of security protection, and in some cases, also define the security metrics. For example, the NIST recommends that the security metrics should be used as a decision making tool when trying to improve the performance of a specific business or situation [157]. The NIST assumes the security metrics for five levels of security. These levels reflect the security policies, procedures, and their applications, compliance testing as well as their effectiveness in the daily operations. Some experts interpret security metrics with respect to their alignment with the organizational goals and existing challenges. The standardization of security metrics would also facilitate the security management of assets, monetary values, intellectual property, and information about and from the customers. Such standards together with the providers' certificate of compliance could create awareness and provide assurances to the customers about the expected security levels [158,159]. The ISO/IEC security standards are summarized in Table XV [160,161].

Table XV
THE ISO/IEC INFORMATION SECURITY STANDARDS

| ISO 27001:2005 ISO/IEC 27001:2013 | Information Security Management System (ISMS) requirements |
|---|---|
| ISO/IEC 27002:2005 | Code of Practice, Guidelines for risk treatment plan |
| ISO/IEC 27006:2007 | International accreditation requirements |
| ISO/IEC  27007 | Guidelines for ISMS auditing |
| ISO/IEC 27011:2008 | ISMS guidelines for telecommunications |
| ISO/IEC 27005:2008 | Guidelines for risk management |
| ISO/IEC 27004:2009 | ISMS measurements |
| ISO/IEC 27003:2010 | ISMS implementation guidance |
| ISO/IEC 18028 series | Network security management |
| ISO/IEC 18043:2006 | Intrusion detection systems |
| ISO/IEC 9797 series | Message authentication codes (MACs) |
| ISO/IEC 15816:2002 | Access control |
| ISO/IEC 18033 series | Encryption systems |
| ISO/IEC 24762:2008 | Business continuity management |
| ISO/IEC 18045:2008 | Evaluation |
| ISO/IEC WD 27007 | Auditing |

In spite of incorporating various security mechanisms, the complexity of modern telecommunication systems makes these systems increasingly more vulnerable to sophisticated threats and attacks. Consequently, highly secure networks resort to providing the Intranet services only, and they are otherwise disconnected from the public Internet. In other cases, safeguarding telecommunication networks against the security attacks requires close collaborations of all the stakeholders and departments within the organization [162]. The proper selection of security metrics can bridge the gap between the information security management, software security and the network security engineering [163]. However, as with any other metrics, it is important to use the security metrics only within the context for which they were developed. Outside this context, the metrics may be ineffective or misleading, and thus, can actually cause the security concerns.

The security metrics can be, in general, categorized with respect to the purpose they are aiming to accomplish. For instance, there are security metrics for the information security, intrusion detection, risk management, impact assessment and so on. Another classification of the security metrics considers a type of attack or threat and its duration. The performance indicators of the used security methods and strategies can be classified as either measuring the performance effectiveness or measuring the performance efficiency [150]. The security metrics related to the performance effectiveness evaluate to what degree the security objectives have been met. This serves as guidance on planning the security measures and objectives. The security metrics related to the performance efficiency measure the proportionality between the intended security objectives and the actual results obtained. They enable to quantify how much effort, time, money and labor were used to fulfill the security objectives. The resources used can be optimized, and the progress of achieving the security objectives can be monitored. The organization can then take a proactive attitude to identify or anticipate any security related problems and solve them in early stages.

Even though making the network secure is costly and it may affect the network services quality and performance, it is often cheaper than the cost associated with having the network compromised [164]. It should be noted that most telecommunication systems evolved with no provisioning for the security, so majority of the current security measures have been added later as extra features. In wired networks, the security threats involve the individual hardware and software components as well as the end-users [147]. The security risks are, generally, higher for more complex hardware components and for closed proprietary software solutions with the online distribution of updates. Furthermore, using the same hardware and software components make the system more vulnerable as the security issues are replicated across the network. So called trusted computing base (TCB) in practice cannot be achieved due to the trade-offs in providing the security while maintaining the system performance, cost and usability [165].

The software-based security technologies have the advantage of being agile, portable, cost effective, and they can be updated as the new vulnerabilities and threat vectors are discovered. However, it is now recommended to embed some security directly into the hardware in order to get protection against the recent most advanced threats [166]. The NIST and several microprocessor firms have defined a hardware root of trust to support the operating systems and applications [167]. The use of hardware security for the Internet of Things (IoT) may overcome problems with the resources constrained devices [168], since cryptographic techniques require an extra computational power and also impose the latency [169]. The biometric systems and smartcards [170] and the hardware firewalls [171] are another example of the recent developments in the hardware based system security. Combining the integrated hardware and software solutions enables to obtain the trustworthy computing spaces [165].

Different malware detection strategies for mobile devices are proposed in [172]. Except malware, personal spyware and grayware are the worst threats for not only Android based mobile devices [173-176], so the regular software updates are very important. Other types of attacks against mobile devices include sniffing, spoofing, spamming and phishing [177]. The security solutions for mobile devices can be implemented at the server side or at the client [177].

The information security including non-repudiation, confidentiality, integrity, and availability (CIA) are fundamental for building the secure computing systems [178]. The communication networks such as the Internet buffer data at the intermediate nodes (routers, switches) which makes them more susceptible to attacks. The network threats are mitigated by the encryption, authentication, authorization, intruder detection, firewalls and other mechanisms [170, 179,180]. The security scopes and objectives as defined by the ITU are given in Table XVI [181]. In the IP-based networks, among the most common security risks are the attacks exploiting malware [179]. The cyber attackers resemble illegitimate profit-seeking entrepreneurs [182] who often seek to commit a financial fraud [183]. An exhaustive review of the security attacks against the IP-based networks is provided in [184]. A comprehensive analysis of the security requirements for the IoT networks can be found in [185]. The authentication and authorization techniques for different IoT applications are considered in [186]. The common attacks against the Internet are summarized in Table XVII [179].

Table XVI
SECURITY OBJECTIVES IN COMMON DATA NETWORKS

| | | |
|---|---|---|
| • Integrity | • Reliability | • Identification |
| • Availability | • Safety | • Recovery |
| • Authentication | • Dependability | • Robustness |
| • Authorization | • Controllability | • Supervision |
| • Non-repudiation | • Audit | • Trustworthiness |
| • Accountability | • Correctness | |

Table XVII
COMMON SECURITY ATTACKS AND DEFENSES IN IP-BASED NETWORKS

| Common attacks | |
|---|---|
| Virus [187] | self-replicating program which propagates by using other files |
| Eavesdropping | unauthorized monitoring and interception of communications |
| Hacking | gaining access by exploiting weaknesses in the target system |
| Worms [188] | self-replicating programs that propagate without the need of a host file (e.g. mass-mailing worms and network-aware worms) |
| Spoofing | source disguised as a known source (IP spoofing can be reduced by address filtering in routers) |
| Denial of Service (DoS) | overloading device and/or network resources, so they become unavailable to users (e.g. TCP handshake can never happen) |
| E-mail bombing and spamming | E-mail bombing is a massive number of identical e-mails sent to one address. E-mail spamming is sending unsolicited e-mails to many users. |
| Phishing | an attempt to obtain sensitive data (e.g. passwords, credit card details) by a third party acting as a legitimate service provider |
| Trojans | malicious computer programs that appear to be benign in order to deliver some payload (e.g. code for remote access and viruses). |
| Side-channel attacks (non-invasive) | Physical attacks in which an adversary tries to exploit physical information leakages, timing and power consumption information, or electromagnetic radiation. In contrast, invasive attacks require depackaging the chip to get the direct access [189], [190]. |
| Fault (active) attacks | Attacks physical access to a cryptographic device and manipulates algorithm execution to inject faulty and retrieve secret material. In contrast, passive attacks observe the devices without disturbing it [191]. |
| IP-based network defenses | |
| Cryptographic systems | methods and tools for protecting information by algorithms and ciphers to encrypt/decrypt plain-text messages and unreadable cipher-texts |
| Firewall | a device or software that controls access to/from the system by filtering incoming and outgoing network traffic |
| Intrusion Detection System (IDS) | a device or software that monitors and automatically alert connections |
| IP Security (IPSec) | security services at network (IP) layer, allowing the choice of security protocols and algorithms [IETF RFC 2401] |
| Secure Socket Layer (SSL) | standard protocol suite for encrypted link between browser and server |

Wireless networks share some common security vulnerabilities with wired networks due to similarity of their protocols [192]. The vulnerabilities, however, differ especially for the threats at the lowest layer of the protocol stack where wireless networks are clearly more vulnerable due to their more open access to the intruders [192,193]. The security of wireless networks can be greatly enhanced by using multiple authentication mechanisms at different protocol layers [192]. Recently, the information-theoretic physical layer security measures have been investigated extensively [194]. Achieving the cyber security in the current as well as the emerging complex telecommunication networks requires contributions from the equipment manufacturers, infrastructure and content providers as well as the support from the government and the policymakers.

The dynamic topology changes and the distributed self-configuration of MANETs poses additional threats to these networks that can be exploited in the routing protocols, in creating selfish nodes, and in wormhole attacks in order to disrupt their normal functionality [195,196]. The security countermeasures may negatively impact the achieved QoS in MANETs [197]. The security of MANETs can be improved by using the intrusion detection techniques, directional antennas, neighbor authentication methods, and by using a secure route delegation [198].

The emerging technologies such as the RFID and the IoT sensors often create new security and privacy risks [185]. Cloud computing have a number of advantages such as transparency, dynamic elasticity, scalability, and economy, but these advantages come at the cost of the security and privacy [199]. There is currently no common security standard for cloud computing, so the cloud vendors implement their own proprietary security solutions and metrics. In addition, an exponential growth of "Big Data" represents similar security and privacy issues which should force the vendors to adopt some security standards [200]. Reference [165] states that, "If security is an attribute of systems and data control (often implemented as technical, physical, and administrative controls), then privacy is the expectation of an individual about maintaining the control of personal information". Confidentiality has similar objectives as privacy, but it focuses on the commitment of keeping information and documents safely, so they can be accessed and shared only by the authorized entities. Furthermore, information sharing in social networking sites may not have direct effects on security, however, the perceived privacy and trust are affected [201].

## 6.  RELIBILITY AND RESILIENCE METRICS

The robustness is defined as "the ability of the system to meet its requirements under a range of representative failure conditions" [202]. It is an important consideration in the design, operation and management of telecommunication systems as it reflects their ability to withstand the changes, for example, in topology and capacity of the links. The robustness can be tested either as a robustness in the network design or as a robustness in the network routing [203]. For the former, the network topology is optimized to offer the desired link capacities despite a range of link failures, malicious attacks or any other such effects. In [204], it is demonstrated that the node connectivity (a graph-theoretical measure) is well suited to measure the network robustness. The effective traffic engineering that achieves the maximum network utilization can also greatly enhance the network robustness in routing [205,206].

The robust networks can be designed to tolerate some maximum level of vulnerability, however, the likelihood of large and potentially disastrous failures should also be considered. The vulnerability of the power grids and of the Internet against random failures and cascading failures has been comprehensively studied in [207]. The cascading failures in such networks can be caused by the targeted as well as random attacks as shown in [208]. Thus, a resilient, robust and fault-tolerant network design against various types of attacks should be considered. The vulnerability of scale-free networks against an overload due to cascading failures (e.g., by exceeding the node capacities) and against fluctuating loads, and large-scale congestion in the scale-free Internet network are explored in [209] and [210], respectively. The schemes against the cascading failures are investigated in [211]. The cost of attacks and the network survivability is evaluated in [212]. The metrics for measuring the network reliability are considered in [213]. Other studies of the cascading failures, vulnerability, robustness and associated network safety issues have been studied in [214-216].

The real-world networks often employ some form of recovery mechanisms to provide a minimum level of network robustness against failures. The recovery mechanisms to cope with the cascading dynamics of random attacks and overload attacks in the weighted scale-free networks are described in [217]. Other network deference and restoration methods to improve the resilience and survivability of telecommunication networks in case of the individual failures are provided in [218]. The quality of recovery (QoR) concept and a comprehensive methodology to account for the availability, quality of a backup path, recovery time, and bandwidth redundancy have been proposed in [219]. The extensive survey [220] presents a variety of methods, procedures and frameworks for the resilience and recovery of networks. However, there seem to be no commonly agreed strategies to define the recovery mechanisms, fault-tolerance and the associated metrics for networks. The resilience differentiation framework is the most promising approach to provide the required resilience levels, and thus, to be also included in the SLAs. It may assist the network operators to stabilize their revenues by reducing the outage events, and to improve the customer satisfaction with the network services.

Similarly to robustness, the network resilience is defined as "the ability of network to provide and maintain an acceptable level of service in face of various faults and challenges to normal operation" [218]. Even though there are conceptual differences between the robustness and the resilience, these terms are frequently used interchangeably. They both express the ability of a system to function properly despite unexpected environmental conditions or internal dysfunctions. These terms correspond to two kinds of system regulations [221]. The functional regulation aims to restore the basic functions of the system whereas the structural regulation restores an equilibrium among the system components. In addition, the robustness is less suitable to be a long-term design strategy while the resilience can be measured only over longer time intervals [222]. Since the robustness and the resilience are somewhat complementary, it is desirable to combine these quantities to enhance the ability of the system to deal with unpredictable abnormal conditions. The robustness is also often used as a synonym for the survivability, reliability or security. The robustness and the resilience are strongly influenced by the system complexity [218]. In particular, the demand for self-organization and autonomy of networks increases their complexity as well as vulnerability, so their robustness and resilience decrease. Other related metrics for the network security, survivability and dependability may also be affected. Defining the robustness target level for a network is non-trivial, since the network robustness depends on many parameters, and it is related to many other metrics as shown in Fig. 5.

The system resilience stems from its trustworthiness and tolerance to external disturbances. The former is a measurable quantity while the latter focuses on the system design to provide services even under the risks. All resilience metrics rely on the robustness, and they can be formally defined as a performance of the control sub-system. The inclusivity of system resilience makes these metrics to be attractive for use with the modern telecommunication systems.

The network robustness metrics can be defined mathematically assuming a graph theory modeling of the network topology [223]. For example, the number of back-up possibilities and the number of alternative paths can be used to quantify the network robustness as shown in [224,225]. In general, the network metrics are graph metrics which describe either the network nodes (e.g., node queue length, and node uptime) and the network edges (e.g., link QoS, and link cost) [226]. The graph theory can be also used to characterize the dynamic processes occurring in a network [227], and to derive the corresponding service performance metrics [228]. In addition, numerous graph based metrics have been devised to predict the survivability and resilience of networks in the presence of various failures and attacks. These metrics are summarized in Table XVIII.

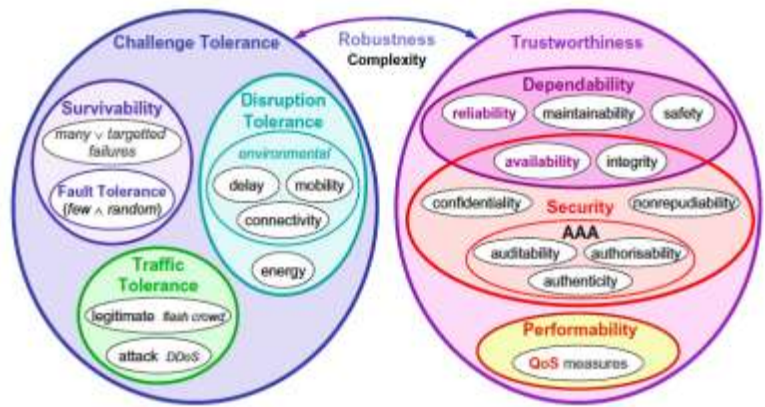

Fig. 5: The resilience and robustness of networks [218].

Table XVIII

COMMON GRAPH MEASURES FOR EVALUATING ROBUSTNESS OF GENERAL NETWORKS

| Metric | Objective |
|---|---|
| Node connectivity [204] | smallest number of node-distinct paths between any two nodes and smallest number of nodes whose removal disconnect the network |
| Heterogeneity [229] | $$\text{Heterogeneity} = \frac{\text{Standard deviation of node degrees}}{\text{Average node degree}}$$ • the smaller, the higher robustness of the network topology |
| Average node degree [230] | connectivity of a network • the larger average node degree, the more robust the network is |
| Symmetry ratio [231] | $$\text{Symmetry} = \frac{\text{\# distinct eigenvalues of adjacency matrix}}{\text{Network diameter}}$$ • symmetric networks perform equally in response to various attacks |
| Clustering coeff. [232] | node clustering (density) is a number of triplets (triangular subgraphs) • network is more robust if its clustering coefficient is higher |
| Average hop-count [233] | the average shortest-paths between all pairs of nodes • smaller number of hops leads to smaller chance of a failed hop |
| Radius [234] | the shortest path of all shortest paths in the network • network is more robust when its radius is small |
| Closeness [235] | measure of centrality as a mean distance from the node to all other nodes |
| Betweenness [235] | To define the number of the shortest paths through a node or link. Using centrality metric to address network robustness can be achieved by evaluating how the network structure changes. |
| Diameter [236] | the longest of all shortest paths between all pairs of nodes • stronger tolerance to attacks for smaller diameter values |
| Average shortest path length (ASPL) [232] | the average of all shortest paths between all node pairs • shorter average path length indicates more robustness |

| Algebraic connectivity [237] | the maximum number of node or link failures a network can accommodate before it becomes disconnected<br><br>• higher value makes network robust against link/node removals |
|---|---|
| Natural connectivity [238] | the redundancy of alternative paths<br><br>• network is robust for higher values of natural connectivity |
| Weighted spectrum (WS) [239] [240] | can be used to identify geographically vulnerable links and nodes |
| Network criticality [241] | the survival of network (its configuration) against topological changes<br><br>• higher robustness when network criticality has smaller value |
| Effective graph resistance [242] | the presence and quality of backup paths between a pair of nodes<br><br>• higher robustness when it is a smaller value |
| Path diversity [233] | number of disjoint alternative paths between two communicating nodes<br><br>Total Path Diversity (TPD) has better accuracy in predicting end-to-end resilience and survivability. |
| Largest eigenvalue [243] [232] | repeated multiplications of adjacency matrix by a nonzero vector<br><br>• higher value of the largest eigenvalue (i.e. small diameter) makes network more robust against link/node removals |
| Assortativity coefficient [232] | the correlation of degrees between nodes<br><br>• higher robustness when the values are close to 0 or 1 |
| Average neighbor connectivity [244] | the average neighbor degree of the average k-degree node<br><br>• higher values indicate higher robustness |

## 7. CONCLUSIONS AND FUTURE RESEARCH

It is clear that the metrics are a key decision tool in the design, analysis, operation and management of telecommunication systems. They are used as KPIs, and they are vital for defining the SLAs. The metrics represent specific models of the system considered. The context in which the metrics are defined and used is important. The context also includes assumptions which are defined either explicitly or more often implicitly about the conditions that have to be satisfied for the metric to be meaningful. The assumptions or conditions can change over time, and the original metric may become misleading. For this reason, some metrics became obsolete, and are phased out or replaced. Likewise, every metric needs to define its measurement conditions. Defining metrics for wireless networks is, in general, more complicated than for wired networks, since the former networks are much more dynamic, operate over unreliable propagation channels, and often involve mobile nodes which can be battery powered. Moreover, the complexity of modern telecommunication systems often requires to assume multiple performance metrics, and the corresponding multi-objective optimization problems.

Technical literature is the greatest source of metrics. However, only a small number of metrics in the literature are being used much more often, so they have become de facto a standard. Most other metrics in the literature are defined and used for specific purpose in a particular scenario. The most important metrics are standardized by various telecommunication institutions, for example, the metrics related to the energy consumption and efficiency. The metric standardization enforces objective comparisons of the products and services from different vendors and service providers. It is possible to define metrics as post-processing of the measured data which is referred to as a reverse modeling of systems, or the required measurement are obtained for given performance metrics which is known as a forward modeling of systems. In general, the metrics and measurements are required to be accurate, valid, feasible, robust, efficient, desirable and viable.

Different stakeholders require and use different types of metrics. The telecommunication systems stakeholders are network infrastructure providers, vendors of telecommunication equipment, service and content providers, regulatory bodies representing the government, and the end-users, and the service and content consumers. Zachman framework capturers the contextual, conceptual, logical, physical, and out-of-context views of the system by different stakeholders. It has been suggested in this paper to use this framework for classifying the metrics.

Many metrics exist or have been proposed for evaluating the performance of telecommunication systems. However, it was observed that the commonly used metrics can be categorized into a relatively small number of groups. The metrics in these groups were surveyed in Sections 2-6. In particular, many energy and power metrics in Section 2 were standardized. The QoS metrics in Section 3 can be further classified as application-based or network-based. They can be measured directly, so they are used as KPIs, or they are included in the SLAs. On the other hand, defining the QoE metrics is more complicated as shown in Section 4. The subjectively assessed QoE metrics are used by the content providers to overcome the limitations of the traditional QoS-aware network designs. The objectively assessed QoE metrics utilize the QoS data to infer the user satisfactions. These metrics can be further subdivided whether they are use the full, partial or no reference. The security metrics were discussed in Section 5. These metrics are gaining significant importance in light of the recent cyber-security breaches. Their standardization is ongoing, however, the telecommunication regulatory bodies currently do not define any specific security protection or intrusion-detection methods. The security metrics can be categorized as information security, intrusion detection, risk management, impact assessment and other metrics. Finally, the reliability and resilience metrics in Section 6 represented the last major category of metrics considered in this paper. These metrics also include the resilience and robustness metrics. They are often derived from the network topology, and are used to forecast the network ability to withstand accidental faults, and targeted damage.

In the following subsection, we outline some possible as well as recommended research directions about developing metrics for telecommunication systems.

## A. FUTURE RESEARCH DIRECTIONS

As pointed out in the introduction, a general theory for systematic, consistent and optimum selection of metrics for a given scenario is missing. This can mean deriving the optimum sensing function for a given system model, or choosing the best metric from some set of feasible metrics. Thus, the metric selection is conditioned on the adopted system model. The optimum metrics likely have to trade-off their accuracy, efficiency, robustness and other attributes. On the other hand, the optimum metrics obtained for each specific case do not allow objective comparisons of products and services from different vendors and service providers. It does not mean that the more universal objective metrics are preferred to the optimum metrics, but these two types of metrics should be used simultaneously, as their purpose is different. More universal metrics are suitable for being standardized, or the metrics can be included in the technology (e.g. 3GPP) standards.

The measurement conditions and procedures represent constraints how the metrics should be used. Despite their importance, they are rarely considered when the metrics are considered. This may be acceptable in technical literature when working with mathematical models of systems. However, in case of the real-world telecommunication networks, the metric values are dependent on the measurement time-scales, where the measurements are taken as well as the network conditions (e.g., traffic load). For instance, many technical papers consider only the energy consumption due to the RF transmissions, whereas the battery powered nodes in real-world networks have to account for the overall energy consumption. It may be useful to consider how the measurements are obtained, communicated and exploited in the network in order to minimize the use of the additional network resources. The robust metrics are less dependent on the specific network conditions, so they may be more reliable and can be used more easily. The robust metrics are also required when we process relatively small number of measurements, or when these measurements are non-ergodic or non-stationary. For network control and management, we need metrics which are predictive, and can support a proactive decision making.

There are many opportunities to improve specific metrics for measuring the system performance, especially as the systems are constantly evolving and new solutions and services and being introduced. For instance, it would be useful to have the QoE metric which can anticipate the user dissatisfaction before they notice a degradation of their experience. The emerging telecommunication systems are becoming more complex which also require to use more complex metrics, or to consider multiple conventional metrics at the same time. However, bundling the metrics can easily lead to various design trade-offs. Moreover, there are no studies to explore how the metrics are used and perceived by different stakeholders of telecommunication systems, since they define and use the metrics independently from each other. Zachman framework may be used to consolidate the stakeholders different views. It would be beneficial to have a general classification schema of the metrics for telecommunication systems which are recommended (not necessarily standardized) for specific use cases, for example, mathematical analysis in technical papers, or for the deployment or operation in some scenarios. A review of implicit assumptions made for common types of metrics can indicate how to use or not to use these metrics. A large body of research can be expected in the area of metrics generated by the artificial intelligence and the machine learning algorithms which can automatically learn the features and extract attributes of the systems.